\documentclass[journal]{IEEEtran}
\usepackage{amsmath}
\usepackage{amssymb}
\usepackage{cite}
\usepackage{graphicx}
\usepackage[dvipsnames,table,xcdraw]{xcolor}
\usepackage{colortbl}
\usepackage{caption}
\usepackage{subcaption}
\usepackage{amsthm}
\usepackage{soul}
\usepackage{hhline}
\usepackage[labelformat=simple]{subcaption}
\usepackage{multirow}

\allowdisplaybreaks

\usepackage[latin1]{inputenc}           

\begin{document}



\title{Non-linear Control of the Power Injected Into a Weak Grid by a Self-Synchronized Inverter}

\author{Sebastian Gomez Jorge$\dag$, Jorge~A.~Solsona$\dag$, Claudio A.  Busada$\dag$, Leire C. Aguirre-Larrayoz$\ddag$, M. Itsaso Martínez$\ddag$ and Gerardo Tapia-Otaegui$\ddag$\\
\thanks{This work was supported in part by the Universidad Nacional del Sur (UNS), in part by Consejo Nacional de Investigaciones Cient\'ificas y T\'ecnicas (CONICET), in part by the Spanish Ministry of Science and Innovation (project code PID2020-115484RA-I00), FEDER Funds, EU, and in part by the Basque Government under research grant IT1644-22.
$\dag$ Authors are with the Instituto de Investigaciones en Ingenier\'ia El\'ectrica (IIIE), Universidad Nacional del Sur (UNS)-CONICET and Dpto. Ing. El\'ectrica y de Computadoras, UNS, Bah\'ia Blanca, Argentina.
$\ddag$ Authors are with the Department of Automatic Control and Systems Engineering, University of the Basque Country UPV/EHU, Faculty of Engineering--Gipuzkoa, Donostia 20018, Spain.
(e-mail:  \mbox{sebastian.gomezjorge@uns.edu.ar}; \mbox{jsolsona@uns.edu.ar}; \mbox{cbusada@uns.edu.ar}; \mbox{leirecheng.aguirre@ehu.eus}; \mbox{mirenitsaso.martinez@ehu.eus}; \mbox{gerardo.tapia@ehu.eus}).}
}


\maketitle

\begin{abstract}
In this work, a non-linear controller designed using non-linear transformation linearization and feedback is proposed for an inverter connected to a weak grid through a single-stage inductive filter. The proposed strategy is self-synchronized, so that it is not necessary to have a voltage sensor at the Point of Common Coupling (PCC). The strategy allows to robustify, in the presence of a weak grid, a strategy that has already been demonstrated to allow a significant reduction in the size of the DC-link capacitor of the converter. For this purpose, a state observer is designed that allows estimating the voltage at the PCC from the measurement of the output inductor current. A start-up controller is also included, which allows synchronization even in the case of system start-up.
Simulation results are presented for different operating cases, including start-up, normal operation, and grid-voltage sags and swells. In all these cases, it is considered that the exact parameters of the grid to which the inverter is connected are unknown.

%
%

\end{abstract}

\begin{IEEEkeywords}
	Grid-tied inverter; power injection; weak grid; non-linear control; auto-synchronization.
\end{IEEEkeywords}

\section{Introduction}
There is an agreement among most nations in the world to develop policies that allow for an increase in energy demand, taking into account environmental aspects. One of the main aspects that must be taken care of is the emission of greenhouse gases into the atmosphere \cite{9705235,10543262}. To achieve this, the two main methodologies used are the capture of gases and the substitution of primary sources of polluting energy for primary sources that avoid or considerably reduce emissions. To increase energy generation and meet environmental standards, the generation of energy using electric generators is in an advanced process of change, where generators based on fossil fuels are being replaced by generators that use non-conventional renewable sources of energy, mainly solar and wind. Most of the time, to be integrated into the electrical grid, generators of this type have an electronic power converter that acts as an interface between the primary source of energy and the grid. This converter consists of a set of switches and some type of filter that combines passive inductive and/or capacitive elements. There are different topologies, and depending on the function that the generator must fulfill in the grid, different control algorithms are designed so that it works as a voltage generator (grid-forming) or a current generator (grid-feeding) \cite{athari2017review,rathnayake2021grid,9552499,9866593,9652035}.

Although there are different topologies and new proposals are continually appearing, a common topology used to make a three-phase generator is to use an inverter connected to a DC link, whose output has an inductive filter. The DC link is often powered by a power source located upstream, and contains a capacitor that stores energy. Many times, this capacitor has a relatively large value to ensure that, in case there are disturbances at the Point of Common Coupling (PCC), there is enough energy to reject them \cite{zolfaghari2022comprehensive}.


When inverters are connected to a strong grid, it has been shown by the authors that non-linear control strategies (see \cite{IFAC2023Gerardo} and \cite{10113842}) allow obtaining a very good performance in presence of large state excursions. Additionally, another advantage introduced by non-linear controllers consists of substantially reducing the DC-link capacitance size \cite{10113842}. In contrast, the classical linear control strategy that uses two decoupled cascade loops (slow external voltage, fast internal current) usually requires a large capacitor.
When using this strategy, the reference for the inner current-loop is constructed considering that the voltage on the DC-link capacitor is slowly varying, forcing the outer control-loop to be slow. This, in turn, results in the slow tracking of the references. Essentially, this is required because linear controllers become unstable when it is necessary to control, with the same linear strategy, the cases in which rapid and wide excursions occur in the states of the generator (i.e., the current of the filter inductance and the voltage in the DC-link capacitor). Additionally, disturbance rejection in the voltage loop is slow. For these reasons, it can be concluded that the great advantage of using a non-linear controller instead of a linear one is that it allows obtaining better system performance in cases where rapid and wide state excursions take place.

However, it should be noted that there are cases where the generator is connected to a weak grid, or even to a grid with completely unknown parameters. If, in this case, the control strategy that was designed considering a strong grid is maintained, it often happens that, for different power injection values, the system becomes unstable. For this reason, many researchers have proposed different strategies to design controllers that guarantee stable system behaviour in the case of connection to a weak grid, but most of them do so by modifying a linear control strategy.
In \cite{8242351} and \cite{9530704}, the stability in the case where the grid has unknown parameters was analyzed. The design of different types of robust controllers is in \cite{xu2017robust,8486047,8490668,wang2020robust}. Other techniques introducing solutions for this problem have been presented in
\cite{9042561,khan2020single,9502732,10609534}.

In a previous work \cite{ARGENCON2024}, considering that the input voltage is provided by a battery, the authors have shown the Steady-State (SS) stability limits for the injection of active and reactive powers guaranteeing that the system remains stable and below nominal current in the presence of a grid with unknown parameters, for the case of both inductive and resistive grids. In addition, they demonstrate the reason why the variables to be controlled exhibit unstable behavior when power is injected into a weak grid, and they have proposed a modification to the control law to make it robust and guarantee transient stability. The main modification to the law consists of feeding back an estimate of the voltage at the PCC, obtained using a notch filter. For the construction of the filter, the measurement of the voltage at the PCC is used, so that it is obviously necessary to employ a sensor to that end. This solution was successfully extended to the case where the power supply to the inverter comes from a capacitive DC-link, where the non-linear controller designed in \cite{10113842} for a strong grid allows to significantly reduce the size of the capacitance in the DC link. The methodology for the design of this non-linear controller is reported in \cite{arXiv.2409.05527}.

One of the objectives of this work is to eliminate the need to measure the PCC voltage and to formulate a control strategy that has the advantage of being non-linear, robust to ensure stability in the presence of a weak grid, and that allows the self-synchronization of the inverter \cite{letterPESobs}. To this end, it is proposed to eliminate the notch filter from the strategy used in \cite{arXiv.2409.05527}, and instead use the PCC voltage estimate obtained through a full-order observer that only requires measuring the current through the output inductor. For safe operation in weak grids, the proposal includes a current limiting algorithm, which will act to protect the inverter against unforeseen variations in the PCC voltage. In addition, to minimize stability problems due to the saturation of the control action that can occur during these transients, Anti-Windup (AW) techniques are also used. Additionally, a start-up algorithm is proposed, and synchronization at system start-up is guaranteed \cite{alqatamin2022nonlinear,liu2022presynchronization}. The proposal is validated through simulation results in various operation conditions.

\section{System Model}
\begin{figure}
	\centering
	\includegraphics[width=\linewidth]{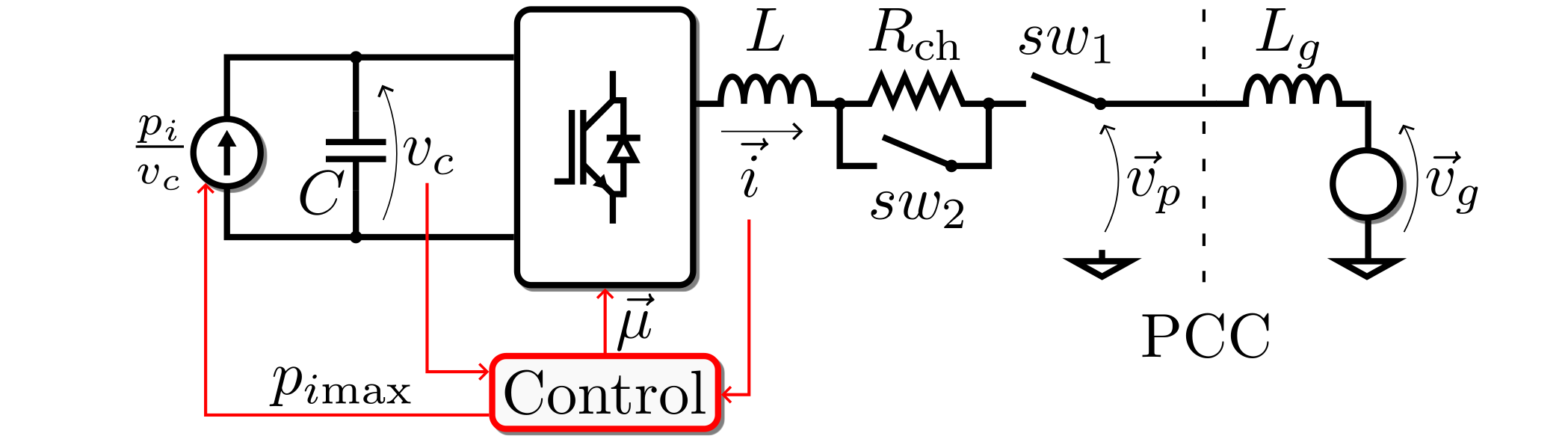}
	\caption{Diagram of the system.}\label{fig:diagrama}
\end{figure}
The diagram of the system under study is shown in Fig. \ref{fig:diagrama}. Here, the three-phase current and voltage variables are represented using complex space-vectors (denoted by $\vec{.}\,$), which are obtained from the power-invariant Clarke transformation. The injected current, PCC voltage and grid voltage are represented by $\vec{i}=i_\alpha+ji_\beta$, $\vec{v}_p=v_{p\alpha}+jv_{p\beta}$ and $\vec{v}_g=v_{g\alpha}+jv_{g\beta}$, respectively, while $\vec{\mu}$ is the modulation index. Switches $sw_1$ and $sw_2$ represent the contactors used to connect the system to the grid, and to short-circuit the DC-link pre-charge resistor, $R_\text{ch}$, respectively. Furthermore, $L$ is the inverter filter inductance, while $L_g$ represents the grid inductance. The DC link is fed by a primary power source $p_i$, which injects current into capacitor $C$, whose voltage is $v_c$. This power source represents, for example, energy from solar panels, batteries, or wind energy, controlled by converters. Although it will be considered that $p_i$ is injected into the DC link asynchronously, since in practice it depends on energy availability, it will also be assumed that it will never exceed the nominal power of the converter. On the other hand, it will be assumed that the converters that provide $p_i$ can receive a signal $p_{i\text{max}}$, sent from the proposed controller (Control block in Fig. \ref{fig:diagrama}), that allows limiting, with a reasonable time delay, the power that they inject into the PCC. This is important in weak grid scenarios, where a decrease in $|\vec{v}_p|$ may result in the inverter not being able to send all the active power entering the DC link to the grid, to avoid exceeding the maximum current. This, in turn, would cause the DC link voltage to grow uncontrollably.

The dynamic model of the system in Fig. \ref{fig:diagrama}, considering $R_\text{ch}$ short-circuited, is given by the following set of equations:
\begin{align}
	L\dot{\vec{i}}&=v_c\vec{\mu}-\vec{v}_p,\label{eq:Ldi}\\
	C\dot{v}_c&=\frac{p_i}{v_c}-\Re\{\vec{\mu}\tilde{\vec{i}}\,\}\label{eq:Cv_c},\\
	(L+L_g)\dot{\vec{i}}&=v_c\vec{\mu}-\vec{v}_g,\label{eq:(L+Lg)di}
\end{align}
where $\tilde{.}$ denotes complex conjugate, and $\Re\{.\}$ is the real part.

\section{Control}
In this work, it is considered that only $\vec{i}$ and $v_c$ are measured. The input power $p_i$ can also be regarded as an available signal sent by the primary source, or it can be estimated by an observer such as the one proposed in \cite{10113842}. On the other hand, $\vec{v}_p$ will be estimated by an observer, which will be described below. In addition, both $L_g$ and the instantaneous value of $\vec{v}_g$ are unknown, although the latter will be considered to be a positive-sequence sinusoidal signal of constant amplitude and angular frequency $\omega$. During the design of the control strategy, it will be assumed that $\vec{v}_p$ is a measured and available signal. Then, in the implementation of the controller, all instances of $\vec{v}_p$ will be replaced by its estimated value, obtained by the observer proposed below. It should be noted that the observer filters the voltage in a similar manner as done by the filter introduced in \cite{arXiv.2409.05527}, but without using the PCC voltage sensor.
%

\subsection{Full-Order Observer of $\vec{v}_p$}
Assuming that in SS
\begin{align}
	\vec{v}_p&=V_pe^{j\omega t},\\
	\dot{V}_p&=0,\\
	\dot{\omega}&=0,
\end{align}
this voltage can be modeled by
\begin{align}
	\dot{\vec{v}}_p&=j\omega \vec{v}_p.\label{eq:dvp}
\end{align}
Therefore, the following full-order observer is proposed to estimate it
\begin{align}
	L\dot{\hat{\vec{i}}}&=v_c\vec{\mu}-\hat{\vec{v}}_p+L\vec{h}_1\vec{\epsilon}_i-(1-sw_2) R_\text{ch}\vec{i},\label{eq:Ldiest}\\
	\dot{\hat{\vec{v}}}_p&=j\omega \hat{\vec{v}}_p+\vec{h}_2\vec{\epsilon}_i,\label{eq:dvpest}
\end{align}
where $\hat{.}$ denotes an estimated signal, $\vec{\epsilon}_i=\vec{i}-\hat{\vec{i}}$, and $\vec{h}_1,$ $\vec{h}_2\in\mathbb{C}$ are constants determining the observer dynamics. The term $(1-sw_2) R_\text{ch}\vec{i}$ in (\ref{eq:Ldiest}), is included to start the observer during DC-link pre-charge, where $sw_2$ represents the contactor command signal in Fig. \ref{fig:diagrama} (1 on, 0 off). This term can be ignored for the analysis that follows.

Subtracting (\ref{eq:Ldiest}) from (\ref{eq:Ldi}) and (\ref{eq:dvpest}) from (\ref{eq:dvp}), the dynamics of the estimation error becomes
\begin{align}
	\begin{bmatrix}
		\dot{\vec{\epsilon}}_i\\
		\dot{\vec{\epsilon}}_{v_p}
	\end{bmatrix}&=
	\underbrace{\begin{bmatrix}
		-\vec{h}_1 & -1/L\\
		-\vec{h}_2 & j\omega
	\end{bmatrix}}_{\boldsymbol{A_o}}
	\begin{bmatrix}
		\vec{\epsilon}_i\\
		\vec{\epsilon}_{v_p}
	\end{bmatrix},\label{eq:Ao}
\end{align}
where $\vec{\epsilon}_{v_p}=\vec{v}_p-\hat{\vec{v}}_p$. From (\ref{eq:Ao}), $\vec{h}_1$ and $\vec{h}_2$ can be easily chosen to place the eigenvalues of $\boldsymbol{A_o}$, thus obtaining the desired dynamics.


\subsection{Current Limiting Algorithm}
In order to protect the inverter, it is proposed to implement a Current Limiting Algorithm (CLA) by means of a Proportional-Integral (PI) current control. This control will only act if the maximum current defined by the designer is exceeded. Otherwise, its effect will be totally cancelled by the Feedback Linearization (FL) control described in Section \ref{sec:FL}.

From (\ref{eq:Ldi}), the following current control algorithm is proposed:
\begin{align}
	\vec{e}_i&=\vec{i}-\vec{i}^*\label{eq:ei}\\
	\vec{u}&=-\vec{k}_p\vec{e}_i-\vec{k}_i\vec{x}_i\label{eq:u}\\
	\vec{\mu}&=(L\vec{u}+\vec{v}_p)/v_c\label{eq:mu corriente}\\
	&\textbf{if }|\vec{\mu}|>\mu_\text{max}\textbf{ then}\label{eq:if mu>mu_max}\\
	&\ \ \vec{\mu}=\mu_\text{max}\vec{\mu}/|\vec{\mu}|;\ \text{sat}_\mu=1\\
	&\textbf{end if}\label{eq:end if mu>mu_max}\\
	\vec{u}&=(v_c\vec{\mu}-\vec{v}_p)/L\label{eq:u sat}\\
	\vec{e}_i&=(\vec{u}+\vec{k}_i\vec{x}_i)/(-\vec{k}_p)\label{eq:ei sat}\\
	\dot{\vec{x}}_i&=\vec{e}_i\label{eq:dxi}
\end{align}
where
\begin{align}
	\vec{u}&=\dot{\vec{i}}\label{eq:u=di}
\end{align}
is an auxiliary control action, $\vec{i}^*$ the current reference, and $\vec{k}_p$, $\vec{k}_i\in\mathbb{C}$ the proportional and integral gains, respectively. The pseudocode (\ref{eq:if mu>mu_max})--(\ref{eq:ei sat}) implements the AW strategy described in \cite{8096954}, which relies on saturating the error $\vec{e}_i$ to prevent the integral state $\vec{x}_i$ from diverging in case the modulation index saturates at $\mu_\text{max}$ ($\mu_\text{max}=1/\sqrt{2}$ when using space-vector modulation). This event is logged in signal $\text{sat}_\mu$, which is zero otherwise. In case this happens, (\ref{eq:u sat}), obtained from (\ref{eq:mu corriente}), computes the value of $\vec{u}$ which replaced in (\ref{eq:u}) gives rise to (\ref{eq:ei sat}), the value of $\vec{e}_i$ keeping the integral state $\vec{x}_i$ at a value such that $|\vec{\mu}|=\mu_\text{max}$.

Replacing (\ref{eq:u}) in (\ref{eq:u=di}) and making use of (\ref{eq:dxi}), the error dynamics become
\begin{align}
	\begin{bmatrix}
		\dot{\vec{e}}_i\\
		\dot{\vec{x}}_i
	\end{bmatrix}&=
	\underbrace{\begin{bmatrix}
		-\vec{k}_p & -\vec{k}_i\\
		1 & 0
	\end{bmatrix}}_{\boldsymbol{A_i}}
	\begin{bmatrix}
		\vec{e}_i\\
		\vec{x}_i
	\end{bmatrix}+
	\begin{bmatrix}
		-\dot{\vec{i}}^*\\
		0
	\end{bmatrix},\label{eq:Ai}
\end{align}
where $\dot{\vec{i}}^*$ can be considered a disturbance, and $\vec{k}_p$ and $\vec{k}_i$ can be easily chosen to locate the eigenvalues of $\boldsymbol{A_i}$ giving rise to the desired dynamics.

\subsection{Feedback Linearization Controller}\label{sec:FL}
To control the voltage in the DC link, $v_c$, and the complex power injected to the PCC, a change of variables will be performed in order to obtain a flat output. This enables controlling the system in the new variables by means of a non-linear transformation that avoids the appearance of zero dynamics. For this, a simplified version of the control proposed in \cite{10113842} will be used (assuming for simplicity that there is no filter resistance). The change of variables is obtained by defining the complex energy $\vec{\xi}_1$ and its time derivative $\vec{\xi}_2$, which represents the complex power balance:
\begin{align}
	\vec{\xi}_1&=\frac{1}{2}(L|\vec{i}|^2+Cv_c^2)+j\eta,\label{eq:xi1}\\
	\vec{\xi}_2&=\dot{\vec{\xi}}_1=p_i-\tilde{\vec{v}}_p\vec{i}=p_i-p+jq,\label{eq:xi2}
\end{align}
where (\ref{eq:Ldi})-(\ref{eq:Cv_c}) has been used to obtain (\ref{eq:xi2}), $p=\Re\{\vec{v}_p\tilde{\vec{i}}\,\}$, $q=\Im\{\vec{v}_p\tilde{\vec{i}}\,\}$ and $\dot{\eta}=q$, with $\Im\{.\}$ the imaginary part. Differentiating (\ref{eq:xi2}) with respect to time, the auxiliary control action $\vec{r}$ of the linearized system is obtained:
\begin{align}
	\vec{r}&=\dot{\vec{\xi}}_2=\dot{p}_i-\tilde{\vec{v}}_p \vec{u}+j\omega \tilde{\vec{v}}_p \vec{i},\label{eq:r}
\end{align}
where (\ref{eq:dvp}) and (\ref{eq:u=di}) have been used. From (\ref{eq:xi1})-(\ref{eq:xi2}), the following algorithm is proposed to control $\vec{\xi}_1$ and $\vec{\xi}_2$:
\begin{align}
	 \vec{r}&=\underbrace{\dot{\vec{\xi}}_2^*-\vec{k}_2\vec{e}_{\xi_2}-\vec{k}_3\vec{x}_\text{fl}}_{\vec{\alpha}}-\vec{k}_1\vec{e}_{\xi_1}\label{eq:r fl1}\\
	\vec{u}&=(\dot{p}_i-\vec{r}+j\omega \tilde{\vec{v}}_p \vec{i})/\tilde{\vec{v}}_p\label{eq:u fl algoritmo}\\
	\vec{i}^*&=(\vec{u}+\vec{k}_i\vec{x}_i)/\vec{k}_p+\vec{i}\label{eq:iref fl}\\
	&\textbf{if }|\vec{i}^*|>i_\text{max}\textbf{ then}\label{eq:if fl}\\
	&\ \ \vec{i}^*=i_\text{max}\vec{i}^*/|\vec{i}^*|;\ \text{sat}_i=1\\
	&\textbf{end if}\label{eq:end if fl}\\
	&\text{Current limiting algorithm (\ref{eq:ei})--(\ref{eq:dxi})}\\
	\vec{r}&=\dot{p}_i-\tilde{\vec{v}}_p \vec{u}+j\omega\tilde{\vec{v}}_p \vec{i}\label{eq:r fl}\\
	\vec{e}_{\xi_1}&=(\vec{r}-\vec{\alpha})/(-\vec{k}_1)\label{eq:aw xi1}\\
	\dot{\vec{x}}_\text{fl}&=\vec{e}_{\xi_1}\label{eq:dxfl}
\end{align}
where $\vec{e}_{\xi_1}=\vec{\xi}_1-\vec{\xi}_1^*$, $\vec{e}_{\xi_2}=\vec{\xi}_2-\vec{\xi}_2^*$, with $\vec{\xi}_1^*$ and $\vec{\xi}_2^*$ references to be defined, $i_\text{max}$ is the current limit, and $\vec{k}_1$, $\vec{k}_2$ and $\vec{k}_3\in\mathbb{C}$ are constants defining the dynamics of the closed-loop system. In this algorithm, (\ref{eq:r fl1}) implements a Full-State Feedback (FSF) controller. Equation (\ref{eq:u fl algoritmo}) is obtained from (\ref{eq:r}) and it is used to compute $\vec{i}^*$ by (\ref{eq:iref fl}) [obtained from (\ref{eq:ei})-(\ref{eq:u})]. In normal operation, where the magnitude of this reference is not saturated by (\ref{eq:if fl})--(\ref{eq:end if fl}), it can be easily verified by replacing (\ref{eq:iref fl}) in algorithm (\ref{eq:ei})--(\ref{eq:dxi}), that the auxiliary control action $\vec{u}$ becomes equal to (\ref{eq:u fl algoritmo}), effectively eliminating the CLA. Furthermore, the integral term $\vec{x}_i$ of the CLA is stable in this case, since replacing (\ref{eq:ei}) and (\ref{eq:iref fl}) in (\ref{eq:dxi}) gives $\dot{\vec{x}}_i=-(\vec{u}+\vec{k}_i\vec{x}_i)/\vec{k}_p$, which is a stable low-pass filter if $\Re\{\vec{k}_i/\vec{k}_p\}>0$. In case the control action is saturated, either by saturation of $\vec{i}^*$ by means of (\ref{eq:if fl})--(\ref{eq:end if fl}) or of $\vec{\mu}$ by means of (\ref{eq:if mu>mu_max})--(\ref{eq:end if mu>mu_max}), the auxiliary control action $\vec{r}$ is recalculated in (\ref{eq:r fl}), since in both cases $\vec{u}$ is modified by the algorithm (\ref{eq:ei})--(\ref{eq:dxi}). This is necessary to implement the AW strategy (\ref{eq:aw xi1}), which relies on saturating the error $\vec{e}_{\xi_1}$ to prevent the integral state $\vec{x}_\text{fl}$ from diverging \cite{8096954}. The event of saturation of $\vec{i}^*$ is logged by signal $\text{sat}_i$, which is zero otherwise.

Let $v_c^*$ and $q^*$, and their time derivatives, be arbitrary reference signals for $v_c$ and $q$, respectively. Then, from (\ref{eq:xi1})-(\ref{eq:xi2}), the references given next are defined:
\begin{align}
	\vec{\xi}_1^*&=\frac{1}{2}(L|\vec{i}^*|^2+C{v_c^*}^2)+j\eta^*,\label{eq:xi1 ref}\\
	\vec{\xi}_2^*&=p_i-p^*+jq^*,\label{eq:xi2 ref}\\
	\dot{\vec{\xi}}_2^*&=\dot{p}_i-\dot{p}^*+j\dot{q}^*,\label{eq:dxi2 ref}
\end{align}
where $\dot{\eta}^*=q^*$, $|\vec{i}^*|^2=({p^*}^2+{q^*}^2)/V_p^2,$ and
\begin{align}
	\dot{p}^*&=\frac{V_p^2(p_i-p^*-C\dot{v}_c^*v_c^*)-L\dot{q}^*q^*}{L(|p^*|+\delta_p)},\label{eq:dpr}
\end{align}
with $\delta_p>0$ an arbitrary constant to avoid division by zero. The derivation of (\ref{eq:dpr}) is detailed in \cite{10113842}, and assumes that $V_p$ is slowly varying. By combining (\ref{eq:xi1})-(\ref{eq:xi2}) with (\ref{eq:xi1 ref})-(\ref{eq:xi2 ref}) the error signals can be constructed directly as follows:
\begin{align}
	\vec{e}_{\xi_1}&=\frac{L}{2}\left(|\vec{i}|^2-\frac{{p^*}^2+{q^*}^2}{V_p^2}\right)+\frac{C}{2}(v_c^2-{v_c^*}^2)+je_\eta,\\
	\vec{e}_{\xi_2}&=-(p-p^*)+j(q-q^*),\\
	\dot{e}_\eta&=q-q^*.\label{eq:e_eta}
\end{align}
The main advantage of constructing these signals directly is that $\eta$ and $\eta^*$ are divergent, while $e_\eta$ is not divergent if the control drives $q$ to $q^*$.

\textbf{Remark 1}: In case the AW algorithm (\ref{eq:if fl})--(\ref{eq:aw xi1}) acts, it is convenient to force the state $e_\eta=0$, since it could be divergent in this case.

\textbf{Remark 2}: It is important to note that, in principle, it is not necessary to know the signal $\dot{p}_i$, which can be considered $\dot{p}_i=0$ for the implementation of the controller. This is because, by replacing (\ref{eq:dxi2 ref}) in (\ref{eq:r fl1}) and the resulting expression in (\ref{eq:u fl algoritmo}), $\dot{p}_i$ cancels out. This is true as long as the saturations (\ref{eq:if mu>mu_max})--(\ref{eq:end if mu>mu_max}) and (\ref{eq:if fl})--(\ref{eq:end if fl}) do not act, since, in case they acted, computation of (\ref{eq:r fl}) would require knowing $\dot{p}_i$. However, the error committed in that case would only affect the AW of $\vec{x}_\text{fl}$ (and only during a transient of $p_i$), which does not influence performance in normal operation.

From (\ref{eq:xi2})--(\ref{eq:r fl1}) and (\ref{eq:dxfl}), the error dynamics result
\begin{align}
	\begin{bmatrix}
		\dot{\vec{e}}_{\xi_1}\\
		\dot{\vec{e}}_{\xi_2}\\
		\dot{\vec{x}}_\text{fl}
	\end{bmatrix}&=
\underbrace{	\begin{bmatrix}
		0 & 1 & 0\\
		-\vec{k}_1 & -\vec{k}_2 & -\vec{k}_3\\
		1 & 0 & 0
	\end{bmatrix}}_{\boldsymbol{A_\text{fl}}}
	\begin{bmatrix}
		\vec{e}_{\xi_1}\\
		\vec{e}_{\xi_2}\\
		\vec{x}_\text{fl}
	\end{bmatrix},\label{eq:Afl}
\end{align}
where $\vec{k}_1$, $\vec{k}_2$ and $\vec{k}_3$ can be easily chosen to locate the eigenvalues of $\boldsymbol{A_\text{fl}}$ leading to the desired dynamics.

\subsection{Droop Control with Input Power Limitation}
An external loop will be added to control $V_p$ to its nominal value. In addition, priority will be given to reactive power injection in order to control $V_p$ without exceeding the current limit, $i_\text{max}$. This implies that a signal must be generated to limit the maximum input power, since, under certain conditions, the converter will not be able to inject this power without the current exceeding $i_\text{max}$ (for example, if $V_p$ is very low). This signal must be sent to the input power source, which needs to be adjusted so as not to exceed the limit established by the reactive power control-loop.

Assuming that the droop control-loop is designed to be slow, the relationship between $V_p$ and $q$ can be modeled using the following steady-state expression \cite{arXiv.2409.05527}:
\begin{align}
	 V_p^2&=X_gq\!+\!\frac{|\vec{v}_g|}{2}\left[|\vec{v}_g|\!+\!\sqrt{|\vec{v}_g|^2\!\!-\!4X_g\!\left(\!\!\frac{X_gp^2}{|\vec{v}_g|^2}\!-\!q\!\!\right)}\right].\label{eq:|vp|^2 ss}
\end{align}
If $p$ is considered as a disturbance that must be compensated by the control loop, it can be neglected in the following analysis. Then, performing a first-order Taylor linearization to (\ref{eq:|vp|^2 ss}) around $q=0$ yields
\begin{align}
	V_p&\simeq\left.\frac{\partial V_p}{\partial q}\right|_{q=0}q=\frac{X_g}{|\vec{v}_g|}q.\label{eq:vp vs q}
\end{align}
Given the dependence of (\ref{eq:vp vs q}) on $X_g$ and $|\vec{v}_g|$, two unknown quantities, a worst-case design will be necessary. Intuitively, the maximum expected value of $X_g$ and the minimum expected value of $|\vec{v}_g|$ should be chosen, such that the gain in (\ref{eq:vp vs q}) is maximum. Any smaller gain should result in a stable closed-loop system. The following PI control algorithm with AW and maximum input power limitation is then proposed:
\begin{align}
	e_{V_p}&=V_p-V_p^*\label{eq:eVp}\\
	q^*&=-g_pe_{V_p}-g_ix_{V_p}\label{eq:q ref}\\
	s_\text{max}&=i_\text{max}V_p\label{eq:smax}\\
	&\textbf{if }|q^*|>s_\text{max}\textbf{ then}\label{eq:if q}\\
	&\ \ q^*=s_\text{max}\,\text{sign}(q^*)\\
	&\textbf{end if}\label{eq:endif q}\\
	p_{i\text{max}}&=\sqrt{s_\text{max}^2-{q^*}^2}\label{eq:pi max}\\
	e_{V_p}&=(q^*+g_ix_{V_p})/(-g_p)\label{eq:aw qr}\\
	\dot{x}_{V_p}&=e_{V_p}\label{eq:dxVp}
\end{align}
where $V_p^*$ is the PCC voltage magnitude reference, and $g_p$, $g_i\in\mathbb{R}$ are constants that allow defining the closed-loop dynamics. In this algorithm, (\ref{eq:smax}) defines the maximum apparent power that can be injected without the current exceeding $i_\text{max}$. If $q^*$ exceeds this magnitude, it is saturated by (\ref{eq:if q})--(\ref{eq:endif q}). Then, (\ref{eq:pi max}) computes the input power limit, which must be sent to the input power source. Finally, in case $q^*$ saturation occurs, (\ref{eq:aw qr}) implements the AW of the integral state $x_{V_p}$, by limiting the error $e_{V_p}$ \cite{8096954}.

Assuming that the PCC voltage control-loop is slow, it turns out that $q\simeq q^*$. Then, combining (\ref{eq:vp vs q})--(\ref{eq:q ref}) and (\ref{eq:dxVp}), the error dynamics becomes
\begin{align}
	\dot{e}_{V_p}&=-\frac{g_iX_g}{|\vec{v}_g|+g_pX_g}e_{V_p}-\frac{|\vec{v}_g|}{|\vec{v}_g|+g_pX_g}\dot{V}_p^*.\label{eq:devp}
\end{align}
Since in general $\dot{V}_p^*=0$, the second term can be neglected. On the other hand, it is clear that the proportional term could have been omitted (since, if $g_p=0$, the dynamics is given by $g_i$). However, this term is included in order to implement the AW (\ref{eq:aw qr}), and, therefore, it can be made arbitrarily small. Assuming that $g_p\ll|\vec{v}_{g\text{min}}|/X_{g\text{max}}$ can be chosen, with $|\vec{v}_{g\text{min}}|$ the minimum expected value of $|\vec{v}_{g}|$ and $X_{g\text{max}}$ the maximum expected value of $X_g$, then (\ref{eq:devp}) can be approximated by
\begin{align}
	\dot{e}_{V_p}&\simeq-g_i\frac{X_g}{|\vec{v}_g|}e_{V_p}.\label{eq:devp aprox}
\end{align}
The integral gain can be chosen to achieve a settling time ---according to the 1\% criterion--- greater than or equal to $\tau$ through
\begin{align}
	g_i&=\frac{4.6}{\tau}\frac{|\vec{v}_{g\text{min}}|}{X_{g\text{max}}}.\label{eq:gi}
\end{align}
\textbf{Remark 3:} Both (\ref{eq:dxi2 ref}) and (\ref{eq:dpr}) require $\dot{q}^*$. Using the approximations made to obtain (\ref{eq:devp aprox}), it turns out that, in general, $g_p\ll g_i$, so this signal could be approximated by (\ref{eq:q ref}) and (\ref{eq:dxVp}) as $\dot{q}^*\simeq-g_ie_{V_p}$. However, since the droop loop is slow, the term $\dot{q}^*$ has no noticeable effect on the transient response of the system, and so it can be approximated by $\dot{q}^*\simeq0$.

\subsection{Start-up Controller}
The system starts with the DC link discharged ($v_c=0$). When closing $sw_1$, $v_c$ will be charged through $R_\text{ch}$, $L$ and the inverter flywheel diodes, reaching a value $v_c\simeq \sqrt{2}V_p$. At that moment, the inverter will be controlled as a variable resistor to bring $v_c$ to its reference value, taking advantage of the fact that the current $\vec{i}$ is limited by $R_\text{ch}$. It is proposed to control the inverter by
\begin{align}
	\vec{\mu}&=-\frac{\kappa(E_c^*-E_c)\vec{i}}{v_c},\label{eq:mu arranque}
\end{align}
where $E_c=0.5Cv_c^2$ is the energy in $C$, $E_c^*=0.5C{v_c^*}^2$ its reference, and $\kappa\in\mathbb{R}^+$ is a gain to be defined. Replacing this result in (\ref{eq:Ldi}) and adding the voltage drop of $R_\text{ch}$ yields
\begin{align}
		L\dot{\vec{i}}&=-[\kappa(E_c^*-E_c)+R_\text{ch}]\vec{i}-\vec{v}_p.\label{eq:Ldi arranque}
\end{align}
Considering that, during the charging of the DC link, $E_c^*-E_c>0$ is true, the total resistance in (\ref{eq:Ldi arranque}) is $\kappa(E_c^*-E_c)+R_\text{ch}\geq R_\text{ch}$. Furthermore, considering that $R_\text{ch}\gg \omega L$, the dynamics in (\ref{eq:Ldi arranque}) is practically resistive, and therefore, the magnitude of the current is bounded by
\begin{align}
	|\vec{i}|\leq\frac{V_b}{R_\text{ch}},\label{eq:cota |i|}
\end{align}
where $V_b$ is the nominal voltage of the grid.
Multiplying (\ref{eq:Cv_c}) by $v_c$, bearing in mind that $p_i=0$ during DC-link charging, and replacing (\ref{eq:mu arranque}) in the resulting expression it turns out that
\begin{align}
	\dot{E}_c&=\kappa|\vec{i}|^2(E_c^*-E_c)\leq\kappa\frac{V_b^2}{R_\text{ch}^2}(E_c^*-E_c),
\end{align}
where (\ref{eq:cota |i|}) has been used to obtain the bound. This implies that the energy dynamics in $C$ is bounded by the response of a first-order filter, and therefore $\kappa$ can be chosen by
\begin{align}
	\kappa&=\frac{4.6R_\text{ch}^2}{\tau_\text{ch}V_b^2},\label{eq:kappa}
\end{align}
where $\tau_\text{ch}$ is the minimum desired settling time.

\begin{figure}
	\centering
	\includegraphics[width=\linewidth]{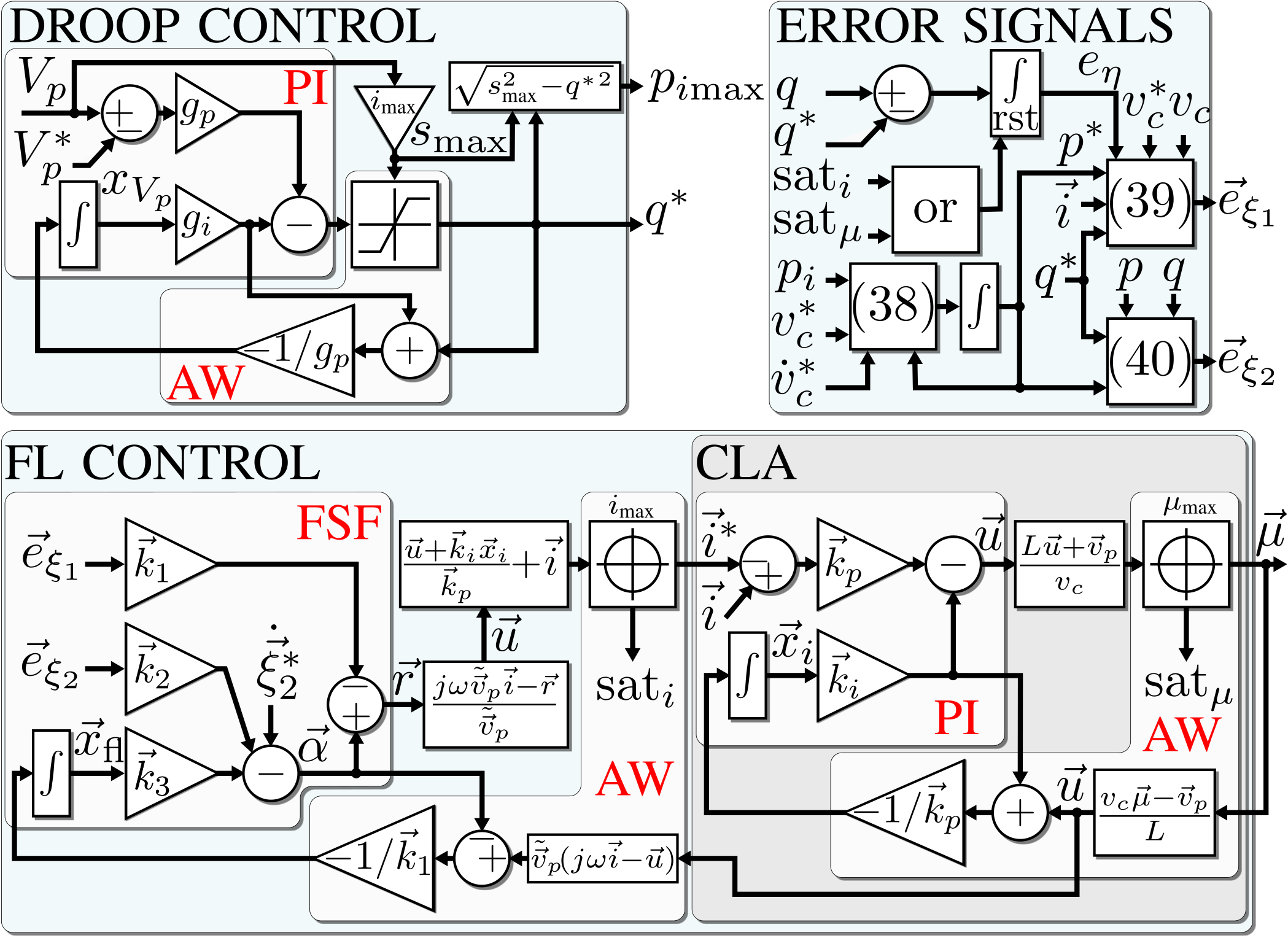}
	\caption{Block diagram of the proposed controller.}\label{fig:control}
\end{figure}
Figure \ref{fig:control} summarizes the proposed controller. The block DROOP CONTROL represents (\ref{eq:eVp})--(\ref{eq:dxVp}), block ERROR SIGNALS represents (\ref{eq:dpr})--(\ref{eq:e_eta}) and block FL CONTROL represents (\ref{eq:r fl1})--(\ref{eq:dxfl}). In the figure, all the instances of $\vec{v}_p$ represent the estimated voltage $\hat{\vec{v}}_p$, obtained by using observer (\ref{eq:Ldiest})--(\ref{eq:dvpest}) (not included in the figure). For simplicity, neither the start-up controller nor the state machine that switches between controllers are included in the figure.

\section{Simulation Results}
\begin{figure}
	\centering
	\includegraphics[width=\linewidth]{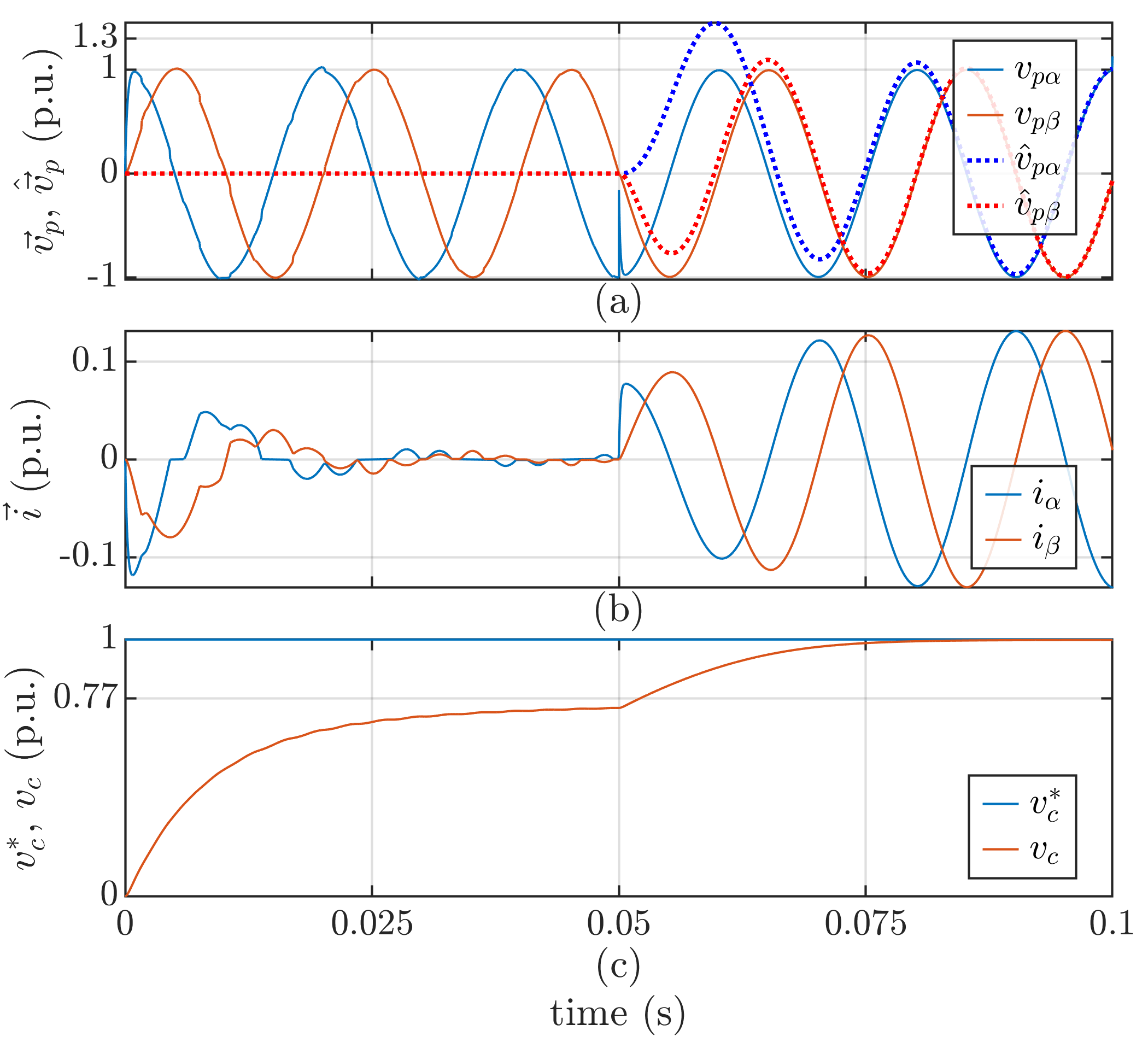}
	\caption{Start-up sequence. (a) $\vec{v}_p$ vs $\hat{\vec{v}}_p$. (b) $\vec{i}$. (c) $v_c^*$ vs $v_c$.}\label{fig:arranque}
\end{figure}
\begin{figure}
	\centering
	\includegraphics[width=\linewidth]{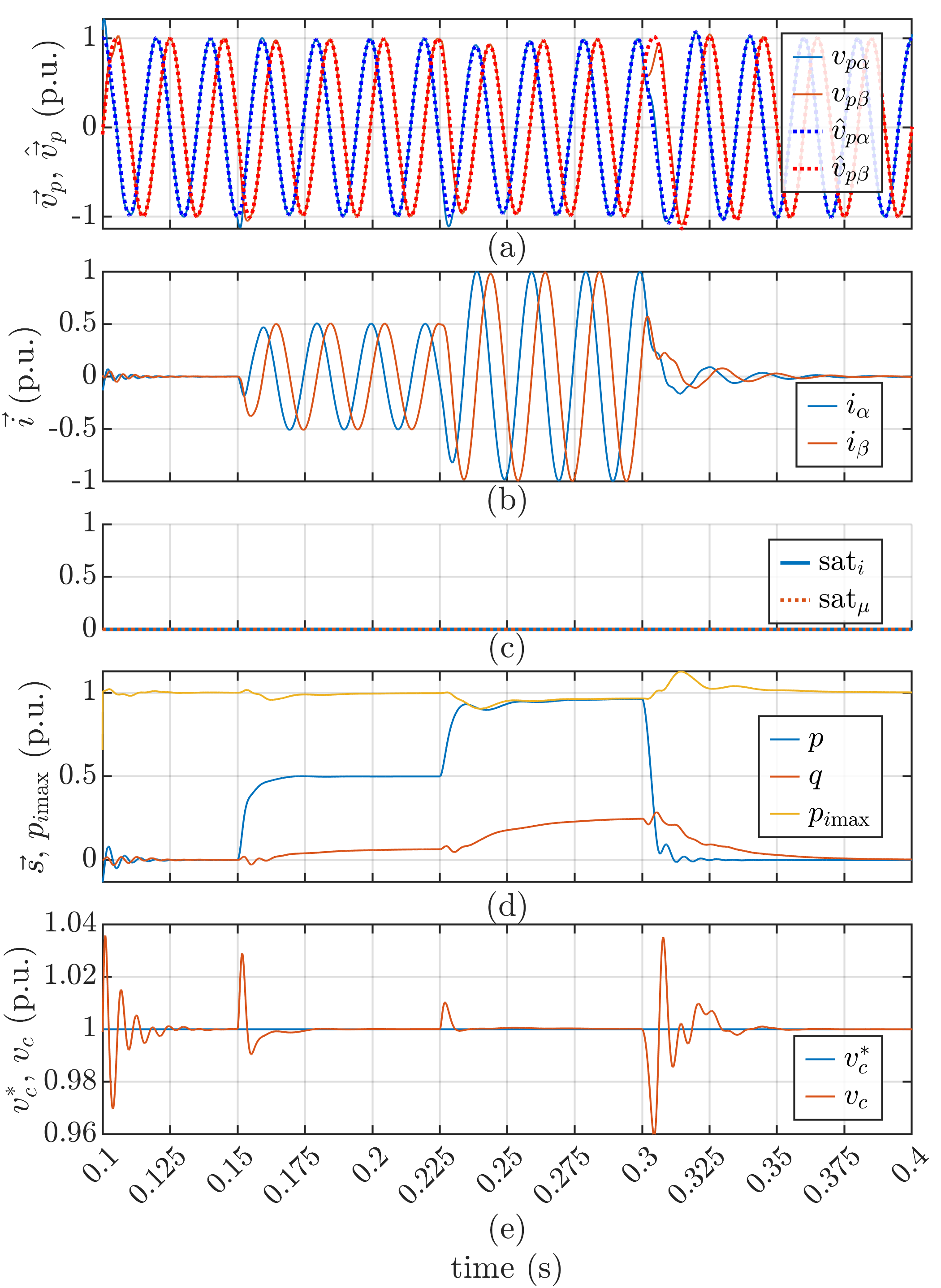}
	\caption{Normal operation. (a) $\vec{v}_p$ vs $\hat{\vec{v}}_p$. (b) $\vec{i}$. (c) Anti-windups for FL control and CLA. (d) $p$, $q$ and $p_{i\text{max}}$. (e) $v_c^*$ vs $v_c$.}\label{fig:potencia}
\end{figure}
\begin{figure}
	\centering
	\includegraphics[width=\linewidth]{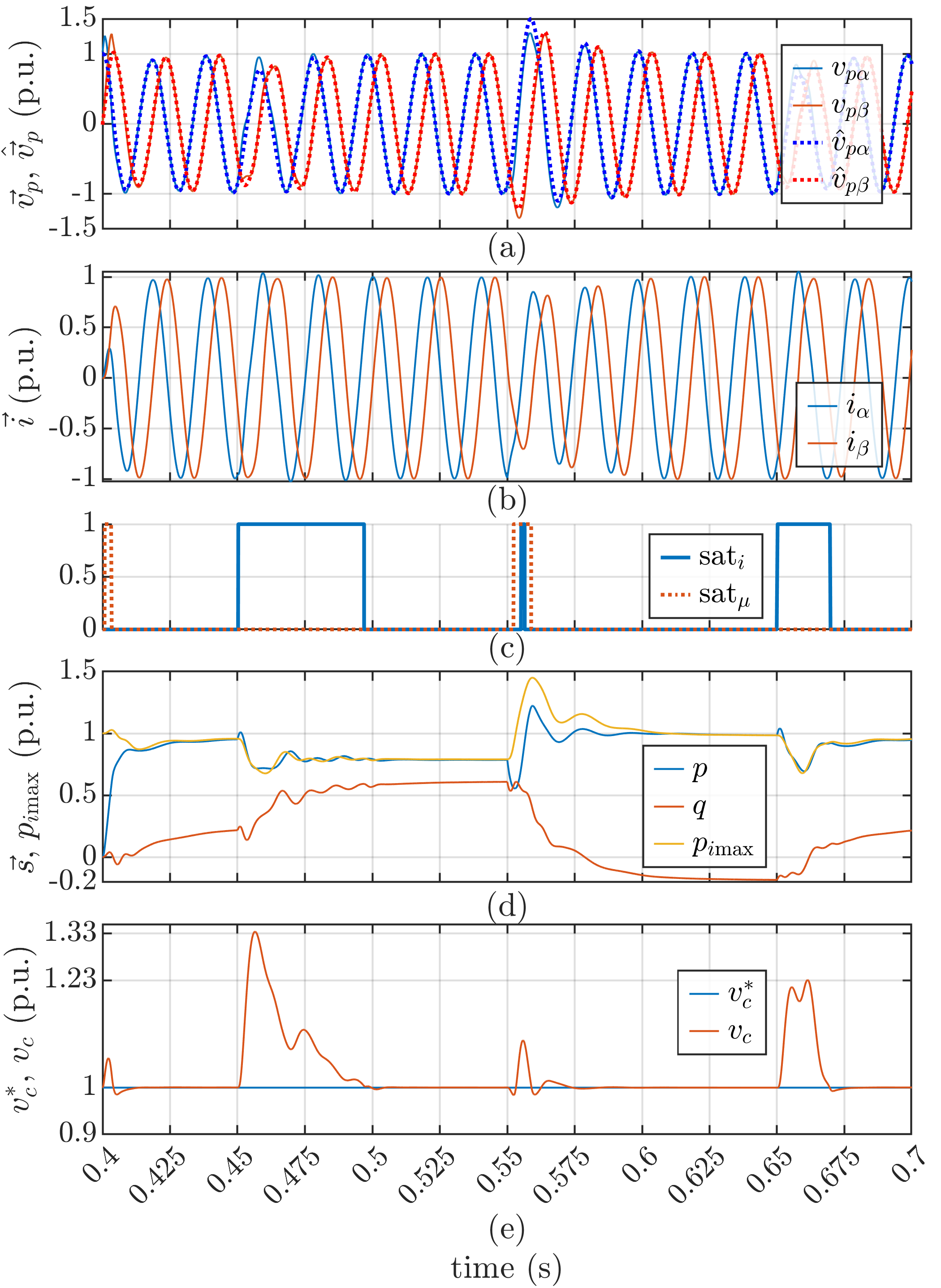}
	\caption{Sag and swell. (a) $\vec{v}_p$ vs $\hat{\vec{v}}_p$. (b) $\vec{i}$. (c) Anti-windups for FL control and CLA. (d) $p$, $q$ and $p_{i\text{max}}$. (e) $v_c^*$ vs $v_c$.}\label{fig:sag swell}
\end{figure}
The system in Fig. \ref{fig:diagrama} will be simulated with the following parameters: nominal power $S_b=2\,$kVA, nominal voltage $V_b=\sqrt{3}\,94\,$V, $I_b=S_b/V_b=7.1\,$A, $Z_b=V_b/I_b=13.25\,\Omega$, $\omega=2\pi50\,$rad/s, $L=2.1\,\text{mH}=0.05Z_b/\omega$, $R_\text{ch}=100\,\Omega=7.54Zb$, $C=48\,\mu\text{F}=1/(5\omega Z_b)$.

The observer gains are chosen by using the $\boldsymbol{A_o}$ defined in (\ref{eq:Ao}) to obtain two single poles with settling times of $5\,$ms and $50\,$ms, respectively. For the CLA, two single poles of $1.5\,$ms and $1\,$ms are chosen, and the gains are obtained from the $\boldsymbol{A_i}$ defined in (\ref{eq:Ai}). For the FL controller, three single poles of $20\,$ms, $1.5\,$ms and $1\,$ms are chosen, and the gains are obtained from the $\boldsymbol{A_\text{fl}}$ defined in (\ref{eq:Afl}). For the droop control, $X_{g\text{max}}=0.8Z_b$ ($L_g=33.7\,$mH), $|\vec{v}_{g\text{min}}|=0.8V_b$ and $\tau=50\,$ms were considered to calculate $g_i$ by using (\ref{eq:gi}), while $g_p=0.01|\vec{v}_{g\text{min}}|/X_{g\text{max}}$ was chosen. Finally, for the start-up control, $\tau_\text{ch}=25\,$ms was chosen and its gain was calculated by using (\ref{eq:kappa}). The voltage reference for the DC link was $v_c^*=300\,\text{V}=1.3\sqrt{2}V_b$, that is, $30\,\%$ above its pre-charge voltage, while the maximum current was $i_\text{max}=I_b$. Additionally, to simulate the dynamics of the set of converters that would provide $p_i$ in a practical application, the response of $p_i$ to the limitation of $p_{i\text{max}}$ is obtained by means of a first-order filter with a settling time of $15\,$ms.

The case of a weak grid with $X_g=0.5Z_b$ ($L_g=21\,$mH) will be simulated. The simulation starts with the inverter in high impedance and $sw_1$ closed, which enables the pre-charge of the DC link through $R_\text{ch}$. Figure \ref{fig:arranque}(a) shows the PCC voltage along with its estimated value obtained from the observer, which is initially disabled, Fig. \ref{fig:arranque}(b) shows the injected current and Fig. \ref{fig:arranque}(c) shows voltage $v_c$ together with its reference (normalized to $v_c^*$). The pre-charge lasts until $t=0.05\,$s, where $v_c$ reaches a value close to $230\,\text{V}=0.77v_c^*$. At that instant, the inverter is started, controlled by the start-up algorithm (\ref{eq:mu arranque}). The observer is also started, considering the last term of (\ref{eq:Ldiest}) to account for the effect of $R_\text{ch}$ on the estimation. As it can be seen in Fig. \ref{fig:arranque}(c), $v_c$ reaches $v_c^*$ within the defined $25\,$ms, and the current in Fig. \ref{fig:arranque}(b) can be seen to have a magnitude less than $0.1325I_b$, as predicted by (\ref{eq:cota |i|}). On the other hand, Fig. \ref{fig:arranque}(a) shows that $\hat{\vec{v}}_p$ converges to $\vec{v}_p$ within $50\,$ms, as designed.

At $t=0.1\,$s, the start-up control is deactivated, the proposed control is activated and $R_\text{ch}$ is short-circuited ($sw_2=1$). The simulation results are shown in Fig. \ref{fig:potencia}, which illustrates the start-up transient lasting approximately $40\,$ms, caused by the increase of $V_p$ due to the short-circuiting of $R_\text{ch}$. The effect on $v_c$ can be seen in Fig. \ref{fig:potencia}(e), where the maximum peak reaches $3.5\%$ of $v_c^*$. Then, at $t=0.15\,$s, the input power $p_i$ rises to $0.5S_b$, as it can be seen in Fig. \ref{fig:potencia}(d), leading to the current rise shown in \ref{fig:potencia}(b). This results in a gradual increase in reactive power, produced by the droop control to bring the PCC voltage to $V_b$. The increment of $p_i$ also results in a transient in $v_c$, but, as it can be seen in Fig. \ref{fig:potencia}(e), it is negligible. From $t=0.225\,$s, $p_i$ increases to $S_b$. However, the drop in $V_p$, which can be seen in Fig. \ref{fig:potencia}(a), reduces the $p_{i\text{max}}$ available in order to keep $|\vec{i}|\leq i_\text{max}$, as seen in Fig. \ref{fig:potencia}(b). The value of $p_{i\text{max}}$ once again tends to $S_b$ as the droop control increases reactive power, bringing $V_p$ back to $V_b$. At $t=0.3\,$s, $p_i$ is reduced to zero, resulting in an increase in $V_p$, which is compensated by the droop control in approximately $75\,$ms. In Fig. \ref{fig:potencia}(c), it is shown that the AW algorithms (\ref{eq:aw xi1}) of the FL control (logged by $\text{sat}_i$) and the CLA (\ref{eq:if mu>mu_max})--(\ref{eq:ei sat}) (logged by $\text{sat}_\mu$) do not act.

To observe a case where the AW strategies act, the simulation continues in Fig. \ref{fig:sag swell}, where the grid voltage undergoes a sag and a swell. In this figure, the power $p_i$ is first increased from zero to $S_b$. This sharp increase to nominal active power results in the $\text{sat}_\mu$ actuation, as shown in Fig. \ref{fig:sag swell}(c), due to the saturation of $\vec{\mu}$ produced by the sharp increase in $V_p$ [see Fig. \ref{fig:sag swell}(a)]. Then, at $t=0.45\,$s, a sag is applied to the grid voltage, leading to $|\vec{v}_g|=0.8V_b$. Figure \ref{fig:sag swell}(c) shows that the CLA AW immediately kicks in ($\text{sat}_i=1$), switching the control operation to current limiting mode. This is due to the reduction in $V_p$ seen in Fig. \ref{fig:sag swell}(a) and the delay with which $p_i$ responds to changes in $p_{i\text{max}}$. The input power that cannot be injected into the PCC increases the magnitude of $v_c$, as seen in Fig. \ref{fig:sag swell}(e). The peak in this case is $33\%$ greater than $v_c^*$, and, if unacceptable, it can be reduced by increasing the value of $C$. As the droop control increases reactive power, the PCC voltage returns to its nominal value, AW condition $\text{sat}_i=1$ is exited, and $v_c$ is brought back to $v_c^*$. Then, at $t=0.55\,$s, a swell in the grid voltage occurs ($|\vec{v}_g|=1.2V_b$). Figure \ref{fig:sag swell}(c) shows that the two AWs act: first $\text{sat}_\mu=1$ due to the saturation of $\vec{\mu}$ (with the increase in $V_p$, $v_c$ is not large enough to control the system), then $\text{sat}_i=1$, due to the increase in $v_c$, which leads the FL control to request $|\vec{i}^*|>i_\text{max}$. In this case, the droop control will inject negative reactive power to reduce $V_p$ back to its nominal value. Finally, at $t=0.65\,$s, the grid voltage returns to its nominal value $|\vec{v}_g|=V_b$. The injection of negative reactive power momentarily reduces $V_p$, which is then recovered by the injection of positive reactive power, showing a transient similar to that observed at $t=0.45\,$s.

The results shown above validate the performance of the proposal, which is capable of providing stable and safe control to the system under very weak grid conditions.

\section{Conclusions}
This work presented a methodology to design a robust and self-synchronized non-linear controller that allows obtaining very good performance when it is necessary to track reference trajectories that vary rapidly, and to reject disturbances that appear at the PCC.

The key to achieving robustness in the presence of a weak grid is to build the non-linear control law based on a non-linear transformation, and to implement the feedback law using an estimate of the voltage at the PCC. To achieve self-synchronization, the estimate of this voltage is obtained by using a state observer built with the measurement of the current flowing through the output inductor.

The simulation results show that the controller presents an excellent performance in different scenarios, managing to satisfy all the properties for which it was designed.

\bibliographystyle{IEEEtran}
\bibliography{bibliosless2}

\end{document}